\documentclass[english,prd,superscriptaddress,nofootinbib,preprintnumbers,twocolumn,showpacs]{revtex4}
\usepackage[utf8]{inputenc}
\usepackage[english]{babel}
\usepackage{amsmath}
\usepackage{amsfonts}
\usepackage{amssymb}
\usepackage{epsfig}
\usepackage{graphics,psfrag,rotating}
\usepackage{graphicx}
\usepackage{dcolumn}
\usepackage{bm}
\usepackage{textcomp}
\usepackage{keyval}
\usepackage{epstopdf}
\usepackage{color}
\usepackage[usenames,dvipsnames,svgnames]{xcolor}
\usepackage[T1]{fontenc}
\usepackage{multirow}
\usepackage{float}

\usepackage{subfigure}

\begin{document}
\title{A $f(R)$-gravity model of the Sunyaev-Zeldovich profile of the Coma cluster 
compatible with {\it Planck} data}
\author{I. De Martino}
\affiliation{Department of Theoretical Physics and History of Science, 
University of the Basque Country UPV/EHU, Faculty of Science
and Technology, Barrio Sarriena s/n, 48940 Leioa, Spain;\\ 
e-mail address: ivan.demartino1983@gmail.com} 

\begin{abstract}
In the weak field limit, analytic $f(R)$ models of gravity introduce
a Yukawa-like correction to the Newtonian gravitational potential.
These models have been widely tested at galactic scales and provide an 
alternative explanation to the dynamics of galaxies without Dark Matter.
We study if the temperature anisotropies due to the thermal Sunyaev-Zeldovich
effect are compatible with these Extended Theories of Gravity. We assume
that the gas is in hydrostatic equilibrium within the modified Newtonian
potential and it is well described by a polytropic equation of state.
We particularize the model for the Coma cluster and the
predicted anisotropies are compared with those measured in the foreground cleaned
maps obtained using the Planck Nominal maps released in 2013. We show that 
the computed $f(R)$ pressure profile fits the data giving rise to competitive constraints of the 
Yukawa scale length $L=(2.19\pm1.02) \rm{\, Mpc}$, and of the deviation parameter $ \delta=-0.48\pm0.22$.
Those are currently the tightest constraints at galaxy cluster scale, and
support the idea  that Extended Theories of Gravity provide 
an alternative explanation to the dynamics of self-gravitating systems 
without requiring Dark Matter. 
\end{abstract}
 
\maketitle
\section{Introduction.}\label{sec:intro}

Astrophysical and cosmological observations clearly indicate that the
Universe has entered into a period of accelerated expansion
\cite{riess, astier, clocc, wmap9, planck15_XIII, bao:blake}. 
The $\Lambda$CDM model accounts for these observations 
by requiring two new energy densities: a Dark Matter (DM) component, characterized 
by a small temperature, interacting only gravitationally with the other 
energy components and a cosmological constant $\Lambda$, equivalent
to a perfect fluid with negative pressure: $p=w\rho$ with an equation of state
parameter $w=-1$. To explain the current period of accelerated expansion only
$w\le -1/3$ is required. This negative pressure fluid is named Dark Energy (DE).
The current values of the energy densities are $\Omega_{DM}\simeq0.26$ and 
$\Omega_\Lambda\simeq 0.68$, in units of the critical 
density, and of the equation of state parameter $w=-1.019^{+0.075}_{-0.080}$,
compatible with a cosmological constant \cite{planck15_XIII}.
Although the effects of DM and DE on large scale are very well 
constrained, the lack of evidence of counterparts at the particle level 
has been interpreted as a break on General Relativity (GR) at galactic, 
extragalactic and cosmological scales and alternative models to GR have
been proposed.  

Generically, Extended Theories of Gravity (ETGs) generalize the 
Hilbert-Einstein Lagrangian by including higher-order curvature invariants   
and minimally or non-minimally coupled terms between scalar fields and geometry. 
The most studied generalization consists in replacing the Ricci
scalar $R$ in the Hilbert-Einstein action with a more general function of the
curvature $f(R)$ (for comprehensive reviews see 
\cite{carroll,sotiriou, defelice, Nojiri2007,Nojiri2011,Capozziello2011}). 
Although many $f(R)$ models can be tested with astrophysical and cosmological 
observations (see \cite{demartino2015a} and reference therein), the exact 
functional form of the gravitational action is still unknown. 
The higher order terms appearing in the Lagrangian can be recast as 
additional scalar fields  by performing a conformal transformation
from the Jordan to the Einstein frame \cite{defelice}. As an example, 
chameleon $f(R)$ models introduce a scalar field non-minimally coupled 
to matter, giving rise to the so called fifth force. These 
models require a screening mechanism to erase 
the effect of the scalar field in high density environments to evade 
the constraints imposed by the Solar System dynamics
\cite{Khoury2004, Mota2004}. Many efforts have been devoted to 
test these models from astrophysical to cosmological scales 
\cite{hu2007-1,hu2007-2, sawicki, song, hu2, hu1, lima, Li21015}. 
For instance, an upper bound on the background amplitude of the chameleon 
field has been found by studying the gravitational interaction on the outskirts 
of galaxy clusters \cite{Terukina2014, Terukina2015, Wilcox2015}.
At the scale of cluster of galaxies, \cite{arnold} demonstrated 
the existence of a degeneracy between the baryonic 
processes and the underlying theory of gravity. 
Although it is always possible to transform the model from the Jordan
to the Einstein frame and vice-versa since they are conformally equivalent, 
nothing can be said {\it a priori} about their physically equivalence that 
must be studied for each specific case \cite{CapFar2010, stabile, Capozziello2011}.

Alternatively, one could make the whole analysis in the Jordan frame and consider the
additional degrees of freedom introduced by the theory of gravity as ``free 
parameters'' that must be constrained using data. Thus,  
those parameters are expected to acquire different values at different scales
in order to pass the Solar system constraints. For instance, analytic $f(R)$ 
models, those expandable in Taylor series around a fixed point $R_0$, i.e.,
\begin{equation}\label{eq:sertay}
f(R)=\sum_{n}\frac{f^n(R_0)}{n!}(R-R_0)^n\simeq
f_0+f'_0R+\frac{f''_0}{2}R^2+\cdots\,.
\end{equation}
give rise to Yukawa-like corrections of the gravitational potential 
in the Newtonian limit \cite{stabile, annalen} 
\begin{equation}
\Phi_{grav}(r)=-\frac{GM}{r} \left(\dfrac{1+\delta e^{-\frac{r}{L}}}{1+\delta}\right)\,,
\label{eq:pot}
\end{equation}
where $\delta$ quantifies the deviation from GR at 
zero order and $L$ is an extra gravitational scale length of the 
self-gravitating object \cite{Capozziello2011}. This new gravitational scale 
is strictly related to the theory: $f(R)$ gravity is a fourth-order theory
and its extra degrees of freedom give rise, in the weak field  limit, 
to a new characteristic scale length (see \cite{hans} for the 
description of the general paradigm on $(2k+2)$-order theories of gravity).  
Those two parameters  are related to the coefficients in 
the Taylor  expansion, eq. \eqref{eq:sertay}, by the relations: 
$\delta=f'_0-1$ and $L=[-{6f''_0}{f'_0}]^{1/2}$ (see Sect. \ref{sec:weakfield} for 
a more detailed discussion). Analytic $f(R)$ models have been shown 
to evade the Solar System constraints \cite{annalen, tsuji, trosi, tartaglia, berry}
and to correctly describe the collapse of self-gravitating systems \cite{jeans},
the emission of gravitational waves \cite{berry, delaurentis2013, delaurentis2015} and 
the dynamics of elliptical and spiral galaxies without requiring DM 
\cite{cardone,napolitano}. Nevertheless, they are poorly tested 
at galaxy cluster scale. At such scale, analytic $f(R)$ models provide a good 
fit to the mass profile of 12 X-ray clusters 
without requiring a DM halo  \cite{salzano}. An equally attractive test
is given by the pressure profile of clusters derived from the Cosmic
Microwave Background (CMB) temperature anisotropies generated by
the thermal Sunyaev-Zeldovich (TSZ) effect  \cite{tsz}.
An upper bound on the parameters of eq.~\eqref{eq:pot} was determined
using a reconstructed map of CMB temperature fluctuations, and computing 
SZ profiles averaged over a sample of 579 galaxy clusters \cite{demartino2014}. 
This led to the bounds $\delta <-0.1$ and $L < 19 \rm{Mpc}$  
at the 95\% confidence level (CL).  

In this paper we demonstrate that the SZ profiles of clusters agree with
the observed profiles when their Intra Cluster gas is in hydrostatic
equilibrium within the potential in eq.~\eqref{eq:pot}. There is
no need to introduce a dominant DM component. We particularize
our analysis for the Coma cluster since is located close to the galactic 
pole where the foreground emission is comparatively low. We
use {\it Planck} 2013 Nominal maps to measure its SZ profile and constrain 
the parameters $(\delta, L)$ of the modified gravitational potential. 
We assume Coma to be spherically symmetric and the Intra Cluster gas to be
in hydrostatic equilibrium. These are good approximations to describe the 
gas distribution and its dynamical state in the intermediate regions where 
the non-thermal pressure is expected to be sub-dominant
\cite{veritas2012, Terukina2014, Terukina2015}.
The outline of the paper is as follows: in Sec.~II, we briefly summarize the 
weak field limit of analytic $f(R)$-gravity and we discuss the 
modified gravitational potential limit at small and large scales,
in Sec.~III, we describe the SZ effect and the model to be tested;
in Sec.~IV, we describe the main observational features of the Coma cluster
and {\it Planck} 2013 Nominal data, summarizing the procedure used 
produced foreground cleaned maps, in Sec.~V we describe our methodology.
Finally, in Sec.~VI we present our results and in Sec.~VII 
we summarize our conclusions.

\section{Yukawa-like correction to the Newtonian potential from the 
weak field limit of $f(R)$-gravity}\label{sec:weakfield}

Our purpose is to test a class of theories of gravity that 
in the weak field limit gives rise to a Yukawa-like correction of 
the Newtonian gravitational potential. Let us first consider the action 
of a $f(R)$-gravity model in the vacuum \cite{carroll, sotiriou, defelice, 
Nojiri2007,Nojiri2011,Capozziello2011} 
\begin{align}\label{eq:FOGaction}
S\,=\,\int d^{4}x\sqrt{-g}f(R).
\end{align}
The corresponding field equations are 
\begin{equation}\label{eq:FEqs}
{\begin{array}{l}
f'(R)R_{\mu\nu}-\frac{1}{2}f(R)g_{\mu\nu}-f'(R)_{;\mu\nu}+g_{\mu\nu}\Box f'(R),
\end{array}}
\end{equation}
and their trace is
\begin{align}
3\Box
f'(R)+f'(R)R-2f(R)\,=0.
\label{eq:TRACE}
\end{align}
In the post-Newtonian limit, one is interested in describing the 
motion of the particles  beyond the 
Newtonian approximation by including  higher order corrections
in the perturbation expansion of the metric. The post-Newtonian limit in a higher 
order theory of gravity introduces correction terms in the Newtonian 
gravitational potential. The  correcting terms depends on the  order of 
the partial differential equations that describe the gravitational field 
\cite{CapFar2010}. For instance, in $f(R)=R^2$ the field equations
are fourth order and in the post-Newtonian limit 
there is a Yukawa-like correction term that modifies
the Newtonian potential \cite{Stelle78}. This correction also appears in any
$f(R)$-model that can be expanded in Taylor series. To illustrate this
point, let us briefly derive the solution of the field equations in the weak 
field limit for a spherically symmetric matter distribution (a detailed 
description of the weak field limit in GR can be found in \cite{weinberg}, 
while its analogue in $f(R)$-gravity was studied in \cite{stabile,CapFar2010}). 
In this case the metric can be written as
\begin{equation}\label{eq:eq12}
d{s^2}={g_{00}}\left({ct,r}\right){c^2}d{t^2}-{g_{11}}\left({ct,r}\right)d{r^2}+ 
{r^2}d{\Omega^2},
\end{equation}
where $d{\Omega^2}$ is the solid angle. 
In order to study the weak field limit, the metric tensor can be 
written as follows \cite{weinberg}
\begin{align}
& {g_{00}}\left( {ct,r} \right) = 1 + g_{00}^{(2)}\left( {ct,r} \right) + g_{00}^{(4)}\left( {ct,r} \right),\label{eq:eq13}\\
& {g_{11}}\left( {ct,r} \right) =  - 1 + g_{11}^{(2)}\left( {ct,r} \right),\label{eq:eq14}\\
& {g_{22}}\left( {ct,r} \right) =  - {r^2},\label{eq:eq15}\\
& {g_{33}}\left( {ct,r} \right) =  - {r^2}\sin {\theta ^2}.\label{eq:eq16}
\end{align}
These expansions are introduced into the field equations \eqref{eq:FEqs} 
to compute
the perturbations at  $\mathcal{O}(0)$, $\mathcal{O}(2)$, and $\mathcal{O}(4)$. 
Particularizing eq. \eqref{eq:sertay} for an analytic $f(R)$ model 
at order zero one finds the condition
\begin{equation}\label{eq:zero_order_condition}
 \frac{f_0}{2}g^{(0)}_{\mu\nu}=0,
\end{equation}
that automatically implies $f_0=0$. Thus, the solutions at higher
orders will not depend on this parameter.
If we now consider the approximation  at second order,
the vacuum field equations can be re-written as 
\begin{align}
& f'_0rR^{(2)}-2f'_0 \partial_r g^{(2)}_{tt}
  +8f''_0\partial_r R^{(2)}-f'_0r\partial_r^2 g^{(2)}_{tt}+ \nonumber \\
&\qquad {} +f''_0rR^{(2)}=0, \label{eq:eq41}\\[2.0mm]
& f'_0rR^{(2)}-2f'_0\partial_r g^{(2)}_{rr}+8f''_0\partial_r R^{(2)}
  -f'_0r\partial_r^2 g^{(2)}_{tt}=0, \label{eq:eq42}\\[2.0mm]
& 2f'_0g^{(2)}_{11}-r[f'_0rR^{(2)}-f'_0 \partial_r g^{(2)}_{tt}
  -f'_0\partial_r g^{(2)}_{rr}+\nonumber \\[0.5mm]
& \qquad {}+4f''_0\partial_r R^{(2)}+4f''_0r\partial_r^2 R^{(2)}]=0, \label{eq:eq43}\\[2.0mm]
& f'_0rR^{(2)}+6f''_0[2\partial_r R^{(2)}+r\partial_r^2 R^{(2)}]=0,\label{eq:eq44}\\[2.0mm]
& 2g^{(2)}_{11}+r[2 \partial_r g^{(2)}_{tt}-rR^{(2)}+2\partial_r g^{(2)}_{rr}
  +r\partial_r^2 g^{(2)}_{tt}]=0. \label{eq:eq45}
\end{align}
These generic expressions can be particularized for a
specific theory selecting the corresponding coefficients $f_i$ in the 
Taylor expansion of eq. \eqref{eq:sertay}. In other words, this system of equations
can be re-written for any $f(R)$-Lagrangian as long as it is expandable in Taylor series.
The solution of eqs. \eqref{eq:eq41}-\eqref{eq:eq45} is \cite{stabile, annalen}
\begin{align}
& g^{(2)}_{tt}=\delta_0-\frac{\Upsilon}{f'_0r}-\frac{\delta_1(t)e^{-r\sqrt{\xi}}}{3\xi
r}+\frac{\delta_2(t)e^{r\sqrt{\xi}}}{6({-\xi)}^{3/2}r}, \\[1.2mm]
& g^{(2)}_{rr}=-\frac{\Upsilon}{f'_0r}+\frac{\delta_1(t)[r\sqrt{\xi}+1]e^{-r\sqrt{\xi}}}{3\xi
r}+ \nonumber \\
 &\qquad \quad {} -\frac{\delta_2(t)[\xi r+\sqrt{\xi}]e^{r\sqrt{\xi}}}{6\xi^2r}, \\[1.2mm]
& R^{(2)}=\frac{\delta_1(t)e^{-r\sqrt{\xi}}}{r}-\frac{\delta_2(t)\sqrt{\xi}e^{r\sqrt{\xi}}}{2\xi},
\end{align}
where $\Upsilon$ is an arbitrary integration constant, the coefficient 
$\xi=\displaystyle-\frac{f'_0}{6f''_0}$ has units of $(length)^{-2}$, and 
$f'_0$ and $f''_0$ are the Taylor coefficients of the Lagrangian;
$\delta_0$ is dimensionless and the time dependent functions $\delta_1(t)$,
$\delta_2(t)$ have dimensions $(length)^{-1}$, $(length)^{-2}$, respectively. 
 Since in the weak field limit  $g_{tt}\,=\,1+2\phi_{grav}\,=\,1+g_{tt}^{(2)}$, 
then $\delta_0=0$ and 
\begin{align}\label{eq:eq49}
 \Phi_{grav}=- \frac{\Upsilon }{2{{f'_0}r}}-\frac{{{\delta _1}(t){e^{ - r\sqrt { - \xi } }}}}{{6\xi r}} + \frac{{{\delta _2}(t){e^{r\sqrt { - \xi } }}}}{{12{{( - \xi )}^{3/2}}r}}.
\end{align}
 When $r\rightarrow\infty$, the potential must go to zero.  Imposing this
condition yields $\delta_2(t)\equiv 0$.  
Finally, to restore GR in the limit $f'_0=1$ and $f''_0=0$, then  $\Upsilon=2GM$.  
Introducing the notation of $\sqrt{-\xi}=1/L$, where now $L$ has units of length,
$\delta_1=-\frac{6GM}{L^2}\frac{\delta}{1+\delta}$, and  
$1+\delta=f'_0$, the eq. \eqref{eq:eq49} can be
rewritten as the modified potential of eq. \eqref{eq:pot}.

\subsection{Small and large scale limits of the modified gravitational potential}
 
The general result of eq. \eqref{eq:pot} implies that the Newtonian potential 
describes the gravitational interaction only in the particular case of the 
Einstein-Hilbert Lagrangian, {\it i.e.} $f(R)=R$. In other words, the parameters 
of the Yukawa-like correction, $(\delta, L)$, represent the deviation of the 
gravitational potential from the standard Newtonian gravity. As  it is
well known, any high order theory of gravity has to evade the Solar System constraints.
Specifically, at small scales ($r\ll L$) the Yukawa term in the gravitational 
potential is not in contradiction with Solar System  observations 
\cite{tsuji, eingorn}.  In fact, in the GR limit
$f'_0=1$ and $f''_0=0$ and $\delta=f'_0-1=0$ and $L=\sqrt{-6f''_0/f'_0}=0$. Thus, 
from eq. \eqref{eq:pot}, the Newtonian potential is restored and no violation 
of the Solar System constraints appears. Since there is no general prescription 
that ensures the physical equivalence of the weak field limit in 
the Jordan and the Einstein frames \cite{Nojiri2006, CapNoj2006_1, CapNoj2006_2}, 
no conformal transformation was made to read these extra degrees of 
freedom as some scalar fields \cite{cap2006, sotiriou2006}. Nevertheless, 
apart of a possible inadequacy in comparing results from the two frames, 
one can re-write any $f(R)$ model as a scalar-tensor theory plus a scalar 
field ($\phi$).  Then, in the small scale limit the mass of the 
scalar field, given by $m^2_\phi=-f'_0/3f''_0=2/L^2$, diverges \cite{cap2008}. 
This divergence corresponds to the well-known chameleon mechanism which 
requires the scalar field to be suppressed 
(at small scales) in high density environments \cite{Khoury2004}.
At large scales, the values $f'_0\neq1$ and $f''_0\neq0$ generate
deviations from Newtonian  gravity, open the possibility to observationally
confirm or rule out these alternative theories of gravity.

Modifications of the Einstein-Hilbert Lagrangian have an effect at all scales. 
As indicated such modifications have to evade the well established 
Solar System constraints. At galaxy scales the corrections to the 
Newtonian gravitational potential can describe systems like spiral and
elliptical galaxies, and galaxy clusters without resorting to DM.  
Detailed analysis at  galaxy scales have been carried out \cite{cardone, napolitano}, 
there is not a definitive answer on whether the assumptions on the underlying
theory of gravity are correct or not.  Our purpose is to test such theories 
using clusters of galaxies, the largest virialized objects in the Universe,
to test modified models of gravity at scales intermediate
between galactic and cosmological scales.
Let us also remark that in the limit $r\gg L$ the gravitational potential 
is that of a point-like mass $M/(1+\delta)$. In this limit, the dynamics
is Newtonian and to explain the structure of self-gravitating systems and
the evolution of large scale structure, DM needs to be introduced as in
the standard cosmological model.

\section{The Sunyaev-Zeldovich cluster profile in $f(R)$-gravity.}\label{sec:profile}

Cluster of galaxies  are the largest virialized objects in the Universe. 
Their Intra-Cluster Medium (ICM) reaches temperatures in the range $T_{\rm e}\sim 1-10$keV.
When CMB photons cross the potential wells of clusters, they gain energy via inverse 
Compton scattering with the hot electrons of the ICM. The CMB
temperature anisotropies  generated by the Sunyaev-Zeldovich effect (SZ)
have two components: the thermal SZ effect (TSZ, \cite{tsz}) due to the thermal
motion of the ICM medium and the kinematic SZ effect (KSZ, \cite{ksz}) due to the 
proper motion of the cluster with respect to the isotropic CMB frame.
The TSZ is usually expressed in terms of the Comptonization parameter $y_c$ as 
\begin{align}
&\frac{\Delta T_{TSZ}(\hat{n})}{T_0}= G(x)y_c =G(x)\frac{k_B\sigma_{\rm T}}{mc^2}
\int_l T_{\rm e}(l)n_{\rm e}(l)dl= \nonumber\\
&= G(\nu)\frac{\sigma_{\rm T}}{mc^2}\int_l P_{\rm e}(l)dl.
\label{eq:y_c}
\end{align}
In this expression $T_0$ is the current value of the CMB black-body 
temperature $T_0=2.725\pm 0.002$K \cite{fixsen}, $G(\nu)$ is
the spectral dependence of the TSZ effect, $\sigma_T$ the Thomson cross section,  
$m_e$ the electron mass, $c$ the speed of light and $k_B$ the Boltzmann constant.
The pressure profile along the line of sight (l.o.s.) $P_e(l)$ is
given by $P_e(l)=n_e(l)T_e(l)$, where $n_e(l)$ and $T_e(l)$ are 
the electron density and electron temperature, respectively.
In {\it the} non relativistic limit ($T_e \approx$ few keV), 
$G(\nu)= \tilde{\nu}{\rm coth}(\tilde{\nu}/2)-4$ where $\tilde{\nu}=h\nu/k_BT_e$ 
is the reduced frequency. To improve the description of the TSZ effect, 
we included relativistic corrections 
in the electron temperature up to fourth order (\cite{itoh1998, nozawa1998, nozawa2006}). 

 To compute the pressure profile of eq. \eqref{eq:y_c} in $f(R)$ gravity, 
we follow the procedure described in \cite{demartino2014}.
We consider the analytic $f(R)$ model given by  eq. \eqref{eq:sertay}, 
and the modified gravitational potential of eq. \eqref{eq:pot}
generalized for an extended spherically symmetric system \cite{salzano}.
Further, the gas is assumed to be in hydrostatic equilibrium within 
the modified potential well of the galaxy cluster 
\begin{equation}\label{eq:HE}
\frac{d\mathbb{P}(r)}{dr}=-\rho(r)\frac{d\Phi_{grav}^{extended}(r)}{dr}
\end{equation}
and it is well described by a polytropic equation of state 
\begin{equation}\label{eq:PES}
\mathbb{P}(r)\propto\rho^\gamma(r)
\end{equation}
The system of equation is closed with the conservation of the mass
\begin{equation}\label{eq:EMC}
\frac{dM(r)}{dr}= 4\pi\rho(r).
\end{equation}
The density $\rho(r)$ refers to the gas density residing in 
the modified potential well of the cluster and does not include 
a DM component.  The pressure profile  
$P(r)=P_c\mathbb{P}(r)$ is given in terms of the two gravitational 
parameters $(\delta, L)$ characterizing the theory, the polytropic index $\gamma$, 
and the central pressure $P_c$. 
The resulting profile will be integrated along the l.o.s. and 
convolved with the antenna beam of the different {\it Planck} channels
to predict the TSZ temperature fluctuations.

\section{Data.} \label{sec:data}

We use the {\it Planck} 2013 Nominal maps\footnote{Data are available at 
http://www.cosmos.esa.int/web/planck} to measure the TSZ temperature 
anisotropies induced by the Coma cluster.  {\it Planck} maps were originally released 
in a Healpix format with resolution $N_{side} = 2048$ \cite{gorski2005}.

\subsection{The Coma Cluster.}\label{sec:Coma}

The nearby Coma Cluster is one of the best studied clusters of galaxies 
because of its richness, degree of symmetry and location near the galactic 
pole.  It is located at redshift $z=0.023$. Its X-ray luminosity and
temperature are $L_X\sim 7.77\times 10^{44}$ erg/s in the $[0.1-2.4]$keV band 
\cite{Reiprich2002} and $T_e=6.9^{+0.1}_{-0.8}$KeV
\cite{Vikhlinin2006}. Recently, the Planck Collaboration \cite{PLANCKX2012} 
determined that the angular size subtended by the $r_{500}$ scale was
$\theta_{500}=48\pm 1$ arcmin, corresponding to $r_{500}\sim 1.314$Mpc
in the concordance model; the associated mass is
$M_{500}\sim6\times10^{14}M_\odot$. Substructure, cooling processes
in the cluster core and departure from the spherical symmetry have
been shown to exist \cite{Coma_xray3, Coma_xray4, Coma_xray5, Coma_xray6, 
Coma_SB,  Coma_XT1, Coma_XT2, Gastaldello, FuscoFemiano}. 
All these components affect the inner- and outer-most regions
of the cluster, but in the intermediate regions the assumptions of 
hydrostatic equilibrium and spherical symmetry hold 
\cite{veritas2012, Terukina2014, Wilcox2015,Terukina2015}. Therefore,
we will restrict our analysis to this intermediate region.

\subsection{Foreground cleaned {\it Planck} Nominal Maps.}\label{sec:cleaning}

{\it Planck} Nominal maps were released in 2013. In addition to the intrinsic CMB
temperature anisotropies, SZ effect and instrumental noise, they contain foreground 
emissions from galactic dust, CO lines, synchrotron radiation and point and 
extended infrared sources. The TSZ effect has a unique dependence with frequency
and it can be reliably distinguished from other components using adequate techniques
to reduced foreground contamination and the cosmological CMB signal (our cleaning
procedure is  extensively described in \cite{demartino2015b}).
We will analyze only the High Frequency Instrument (HFI) data. 
This instrument operates at frequencies 100-857 GHz, 
with angular resolutions $\theta_{FWHM}\leq 10$ arcminutes. It has better
angular resolution and lower instrument noise 
than the Low Frequency Instrument (30-70 GHz). We will measure the TSZ 
cluster profile at 100, 143, and 353 GHz channels, since
after cleaning the data at 545 GHz is still dominated by residual foreground emission.
In those channels, Coma pressure profile has been measured from  5 to 100 arcminutes.
At 217 GHz the TSZ is greatly reduced and provides no relevant information.
The details on {\it Planck} data processing are deferred to App. \ref{app:data}.

\section{Methodology.}\label{MCMC}

To constraint extended theories of gravity using the pressure profile of the Coma
cluster, we first measure the average TSZ emission 
($\delta \bar{T}(\nu_k, \theta_i)/T_0$) over disc/rings
out to $\sim 100$ arcminutes from center of the Coma cluster (see App. \ref{app:data}). 
The model prediction is computed at the same apertures
($\delta \bar{T}({\bf p}, \nu_k, \theta_i)/T_0$) (see Sec. \ref{sec:profile}), and
fit to the data. We compute the likelihood $-2\log{\cal L}=\chi^2({ \bf p}, \nu_k)$ as
\begin{equation}
-2\log{\cal L}=\chi^2 ({ \bf p}, \nu_k)=\Sigma_{i,j=0}^{N} 
\Delta \bar{T}_{ki}({\bf p})C_{ij}^{-1}
\Delta \bar{T}_{kj}({\bf p}),
\label{eq:chi}
\end{equation}
where $\Delta \bar{T}_{ki}({\bf p})\equiv\dfrac{\delta \bar{T}({\bf p}, 
\nu_k, \theta_i)}{T_0}-\dfrac{\delta \bar{T}(\nu_k, \theta_i)}{T_0}$,
$N=22$ is the number of data points, ${\bf p}=[P_c, \gamma, \delta, L]$ are 
the parameters of the model, $\nu_k$ denotes each {\it Planck} channel and 
$C_{ij}$ is the correlation matrix (see eq. \eqref{eq:Cij}).
When computing the likelihood we neglect the error on the CMB blackbody  
temperature $T_0$ and the uncertainty on the angular scale that corresponds
to each data point since they are both negligible with respect to the error 
on the TSZ temperature anisotropies.

\subsection{Monte Carlo Markov Chain sampling method}\label{sec:MCMC}

We constraint the $f(R)$ model parameters using a Monte Carlo Markov Chain 
(MCMC). We employ the Metropolis-Hastings \cite{Metropolis1953, Hastings1970} sampling 
algorithm and we use the Gelman-Rubin criteria to test  the mixing and
convergence of our runs  \cite{Gelman1992}.  We run four independent chains 
starts at a random point of the parameter space
and contains at least 40,000 steps. The step size is adapted in order
to reach an optimal acceptance rate between 20\% and 50\% \cite{Gelman1996, Roberts1997}.
If the step size is too small the acceptance rate will be too high 
($\ge 50\%$) resulting in poor mixing and 
if the step is too large then the acceptance rate will be 
small ($\le 20\%$) and the chain will converge slowly. 

\begin{table}
\begin{center}
\begin{tabular}{|cc|}
\hline
 {\bf Parameter} & {\bf Priors} \\
 \hline
 $P_c /[10^{-2}$ cm$^{-3}$ keV]  & $[0.0, 3.0]$\\
 $\delta$ & $[-0.999, 1.0]$\\
 $L$/[Mpc] & $[0.01, 20]$\\
 $\gamma$ & $[1.0, 5/3]$\\
 \hline
\end{tabular}
\caption{Parameter space explored by the MCMC}\label{tab:priors}
\end{center}
\end{table}
The parameter space explored by our MCMC is given in Table \ref{tab:priors}.
Intervals are chosen on physical grounds: the central pressure $P_c$ 
must be positive; in the central pixel we measure $y_C\simeq 150\mu$K that 
corresponds to $\sim1.37 \times10^{-2}$ cm$^{-3}$ keV, thus 
$[0.0-3.0]\times10^{-2}$ cm$^{-3}$keV represents a  suitable range;
ICM models having the polytropic index $\gamma > 5/3 $ 
are convectively unstable \cite{Sarazin1988}, therefore $\gamma$ is varied between 
the isothermal and the adiabatic limits
$\gamma=[1.0, 5/3]$; the gravitational potential diverges at
$\delta=-1$ and becomes repulsive at $\delta<-1$, 
following \cite{demartino2014} we set $\delta=[-0.999,1.0]$. This 
range also includes $\delta=0$ for which the Yukawa term disappears and 
the cluster gravitational potential is Newtonian,  shallower than
in the concordance model since it does not contain DM, being generated only by the
baryonic component.  The gravitational scale length $L$ varies
between two limiting scales: from galaxy scales \cite{napolitano} to the 
mean cluster separation scale, $L=[0.01,20]$ Mpc.  Finally, we merged the 
four chains and used the marginalized distributions to compute the best fit 
parameters and their $1\sigma$ errors bars.

\section{Results and Discussion.}

To test for consistency, we performed two different analyses. First, MCMC
are run independently fitting separately the data at each frequency. 
Second we computed the joint 
likelihood ${\cal L}({\bf p})=\Pi_k{\cal L}({\bf p}, \nu_k)$. We found that 
the best fit parameters at the three frequencies are consistent with 
each other and with the results obtained from the join distribution, 
indicating that our cleaning procedure does not distort
the TSZ emission of the cluster and that the data  at each frequency
only differ in the level of the remaining foreground residuals.
Due to this internal consistency  hereafter we will only quote the results 
derived from the joint analysis.

\begin{figure}
\epsfxsize=1.0\columnwidth \epsfbox{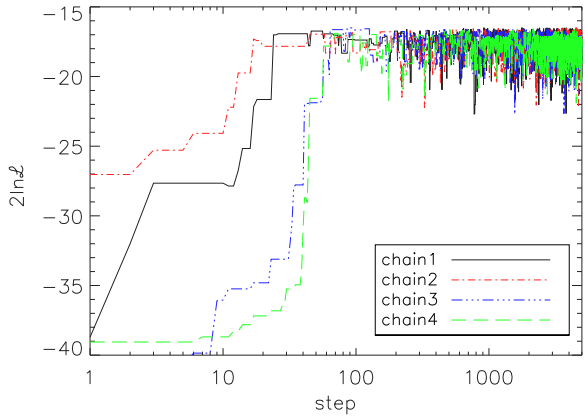}
\caption{Trace plot of the Likelihood values of the four MCMC. 
Only the first 5,000 steps have been represented 
to better visualize the burn-in phase.}\label{fig2}
\end{figure}

Fig.~\ref{fig2} illustrates the  MCMC convergence after an initial burn-in phase 
($\lesssim 100$ steps). We represent the trace plot of likelihood values of the first 
five thousand steps of each chain. To ensure good mixing we use the Gelman-Rubin 
criteria requiring the ratio of the variances in the target distribution 
to be $\mathcal{R}<1.03$ (see Sec. 3.2 of \cite{Verde2003} for definitions). 
In our chains we found $\mathcal{R}=1.007$.  

\begin{figure*}
\epsfxsize=1\textwidth \epsfbox{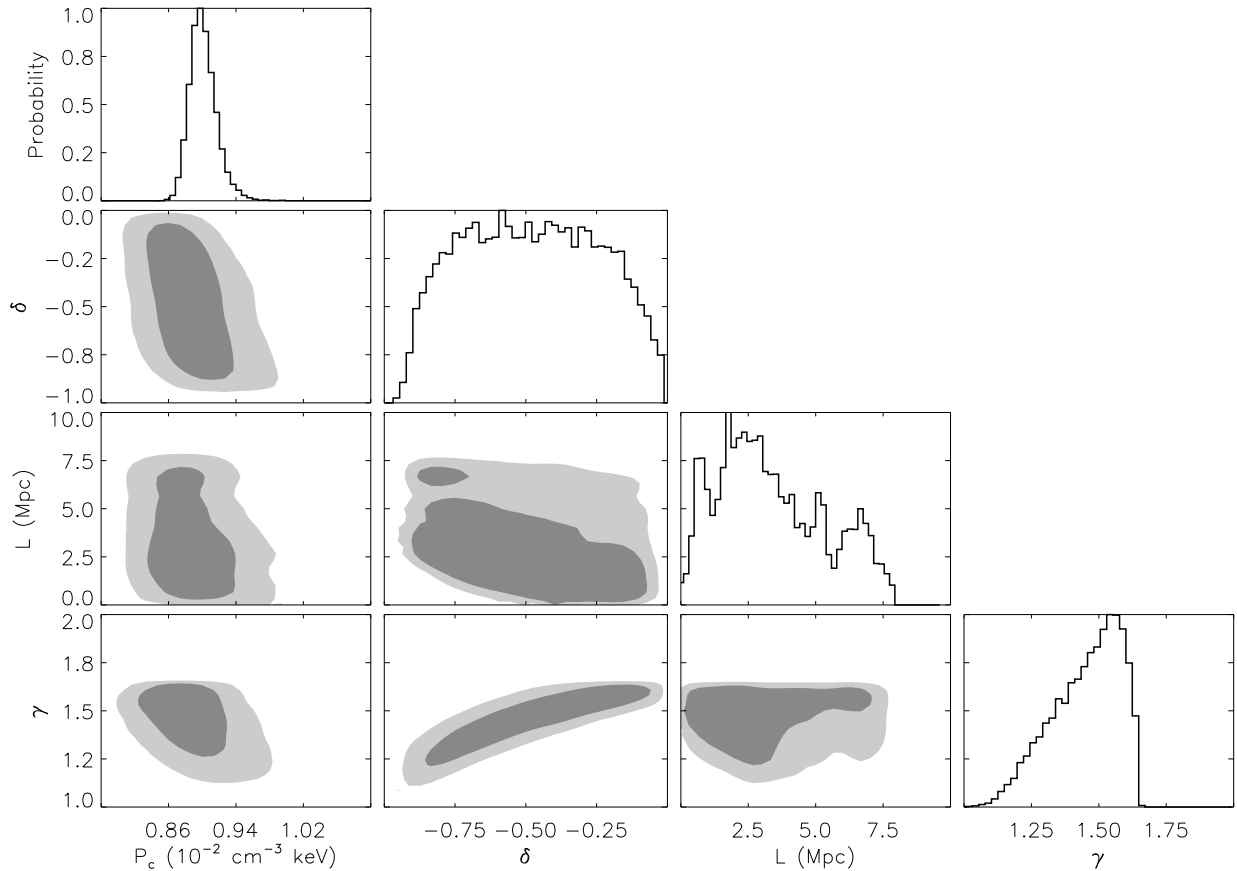}
\caption{2D marginalized contours of the model parameters 
$(P_c, \delta, L, \gamma)$  obtained from the MCMC analysis.
For each pair of parameters the 68\% (dark gray) and 95\% CL (light gray) are shown.
In each row, the
right panels represent the marginalized likelihood distributions.}\label{fig3}
\end{figure*}

In Fig.~\ref{fig3}, we represent the 2D probability contours at
the 68\% CL (dark gray) and  
95\% CL  (light gray) for pairs of parameters and the 1D probability distribution 
of each parameter. Compared with \cite{demartino2014}, the 2D contours are closed 
allowing us to derive new constraints on $(\delta, L)$; these 
results are summarized in Table~\ref{tab:results}
\begin{table}
\begin{center}
\begin{tabular}{|l|c|}
\hline
 {\bf Parameter} & {\bf Results} \\
 \hline
 &\\[-0.2cm]
 $P_c (10^{-2}$ cm$^{-3}$ keV)  & $0.90\pm0.04$\\[0.2cm]
 $\delta$ & $-0.48\pm0.22$\\[0.2cm]
 $L$ (Mpc) & $2.19\pm1.02$\\[0.2cm]
 $\gamma$ & $1.44^{+0.10}_{-0.17}$\\
 \hline
\end{tabular}
\caption{Results from the MCMC.}\label{tab:results}
\end{center}
\end{table}

The table summarizes two important results:
for the first time the central amplitude of the TSZ profile in Coma cluster has been 
constrained in $f(R)$ gravity and $\delta\ne 0$ at the 95\% CL;  
 this latter result clearly shows that the data is compatible with f(R)-gravity plus baryons,
and could be interpreted as statistical evidence in favor of modified gravity and 
against GR plus DM.
The value $L=0$ is also ruled out at the $95\%$ of CL. This limit
corresponds to a Newtonian gravitational potential generated 
by a mass $M'=M/(1+\delta)$. Since the data favors models with $\delta<0$,  $M'\gg M$ 
is analogue to the field generated by a cluster that contains a large fraction of 
DM distributed like the baryonic gas. 
To conclude, \emph{to explain the pressure profile of the COMA cluster,
either DM or a modified theory of gravity is required.}
We found that the preferred value of the polytropic index 
is $\gamma=1.44^{+0.10}_{-0.17}$. The best fit value is compatible at 
the 1.5$\sigma$ level with $\gamma\sim1.2$, the value preferred by observations 
and numerical simulations \cite{Ostriker2005, Ascasibar2006, Bode2009, capelo2012}.
Since the polytropic index is determined by 
physical processes that drive the cluster collapse and its subsequent relaxation 
\cite{Bertschinger1985}, our results support the idea that analytic $f(R)$ models 
without a dominant DM component can explain the structure of galaxy clusters
as the $\Lambda$CDM model. 

\begin{figure}
\epsfxsize=1.0\columnwidth \epsfbox{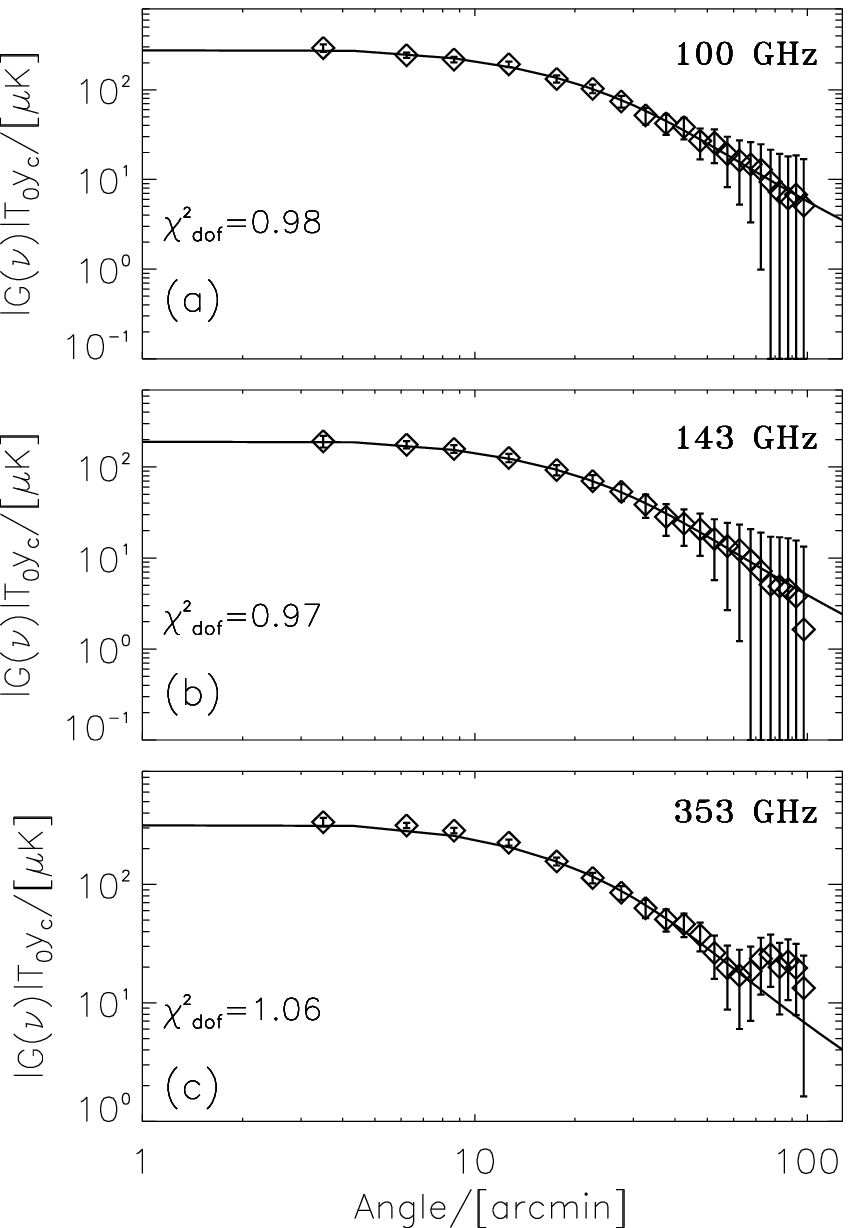}
\caption{Predicted and measured TSZ profile of the Coma cluster at different 
frequencies. For each channel, the $f(R)$ best fit model has been convolved 
with the antenna beam.}\label{fig4}
\end{figure}

To illustrate the quality of the fit to the data by the model, in  
Fig. \ref{fig4} we plot the data (diamonds) with their associated error bars and 
the best fit  model (solid line). Panels (a-c) correspond to the three different
frequencies. The $\chi^2$ per d.o.f is given in each panel. The plot shows a 
slight departure from the model in the outskirt of the cluster, specially
evident at 353 GHz. Since our clean
patches are generated centered on the cluster, the outer parts of the
cluster profile is the region most affected by foreground residuals. This is particularly
true at 353 GHz where we estimate a residual of $\sim 5\mu$K compared with
$\sim 1\mu$K at the other two frequencies. However, this departure could
be due to physical reasons such as the gas not being in hydrostatic equilibrium,
departures from spherical symmetry and the presence of substructure.
 In order to study the impact of such phenomena one should include 
a non-thermal term in the pressure. For that, either N-body simulations of each
specific $f(R)$ model  or a scaling relation  to estimate such
non-thermal component are needed. Since the latter has been 
achieved using hydrodynamical N-body simulations in $\Lambda$CDM, its use could
bias the results forcing the model to mimic the DM in the outskirt masking any effect
due to the modified theory of gravity.
Finally, it has been shown in  \cite{Terukina2014, Terukina2015} that such contributions 
are negligible when, as in this work, the analysis is restricted to the intermediate regions of the cluster.

The resolution of our foreground clean maps is 10 arcminutes and the data do not probe the 
innermost region where shock heating, turbulence, cooling flows and other physical
processes can deviate the dynamical state of the gas from hydrostatic equilibrium. Also, we did 
not stack the cluster profile  outside $\theta > 100$ arcminutes since dust residuals dominate over
the TSZ emission. To test the theory of gravity using data from the cluster core would require a
good understanding of the physical phenomena to separate the physical effects associated with 
the dynamical state of the gas from those associated with the modified theory of gravity and 
an extensive study with numerical hydro-codes would be needed to compute the pressure profile. 
In any case, the next-generation of full sky CMB missions such as COrE/PRISM \cite{PRISM} will resolve 
the core, opening to the possibility to properly study the coupling/degeneracies between the 
underlying theory of gravity and the baryonic processes.

\section{Conclusions.}

Cluster of galaxies have been widely used to constrain  chameleon $f(R)$ models 
(\cite{Terukina2014, Wilcox2015, Li21015}), the Galileon gravity model \cite{Terukina2015},
the K-mouflage modification of gravity \cite{Brax2015} and  
$f(\chi)$ gravity models \cite{bernal2015}.
We have constrained a particular class of ETGs using the  
TSZ temperature anisotropies due to the Coma cluster. With respect 
to previous works, in our model we introduce a Yukawa-like correction to the
Newtonian potential in the weak field limit of an
analytic $f(R)$ model. To test such model, we have produced
foreground cleaned patches of the Coma cluster from the 2013 data release
of {\it Planck} Nominal maps. The measured TSZ profile was used to constrain
the model parameters using an MCMC algorithm assuming that the gas 
was in hydrostatic equilibrium  within the potential well
and the physical state of the ICM was well described by a polytropic equation of state.
The correctness of these hypotheses can only be tested using
hydrodynamical simulations. However since ETGs have extra degrees of freedom compared
to GR, each specific Lagrangian would require its own set of
simulations. Lacking this information we could not study the effect of
a possible departure of hydrostatic equilibrium as well as possible degeneracies 
between baryonic processes and the underlying theory of gravity.

Even though we have studied a single cluster, the marginalized likelihood
functions displayed closed contours. This analysis provided a more sensitive
statistic and allows us to improved our earlier results \cite{demartino2014}. 
The best fit values of the model parameters are summarized in Table~\ref{tab:results}.
We confirm that the value $\delta \simeq 0$ and $L=0$ are ruled at the 95\% CL,
demonstrating that the dynamical state of the Coma cluster gas is
either in equilibrium of the gravitational field of a DM halo or the gravitational
field is modified by a Yukawa-type correction.  This latter idea finds 
additional support since the 
best fit value of the polytropic index $\gamma=1.44^{+0.10}_{-0.17}$ is compatible
with observations and numerical simulations.

\section{Acknowledgments}

IDM was supported by University of the Basque Country UPV/EHU under the program
"Convocatoria de contrataci\'{o}n para la especializaci\'{o}n de personal 
investigador doctor en la UPV/EHU 2015", and by the Spanish Ministerio de
Econom\'{\i}a y Competitividad through grant FIS2014-57956-P (comprising FEDER funds).
IDM warmly thank F. Atrio-Barandela and M. de Laurentis for useful comments and discussions.

\appendix

\section{}\label{app:data}

In this Appendix we briefly describe the method used  to clean
{\it Planck } 2013 Nominal maps and the errors on the measured TSZ profile.
More details are given in \cite{demartino2015b}.  Briefly, 

\textbullet \, all {\it Planck} channels are downgraded to a common 10 arcmin
resolution assuming a Gaussian beam for; \\

\textbullet \, at each frequency, the cosmological CMB and KSZ signals are removed by 
subtracting the LGMCA CMB template of \cite{bobin2013, bobin2014};\\

\textbullet \, CO emission at 100 and 217~GHz is removed by subtracting the CO 
Type 2 maps provided by the Planck Collaboration \cite{planck13_XIII}, \\

\textbullet \, point sources and foreground residuals close to the
Galactic Plane are excised with the PCCS-SZ-Union mask 
\cite{planck13_XVIII, planck13_XXIX}; \\

\textbullet \, finally, the thermal dust emission is removed using the 857~GHz 
channel as a dust-template \cite{planck13_XI, planck13_XII}, following the method 
described in \cite{diego2002}. \\

\begin{figure*}
\epsfxsize=1.0\textwidth \epsfbox{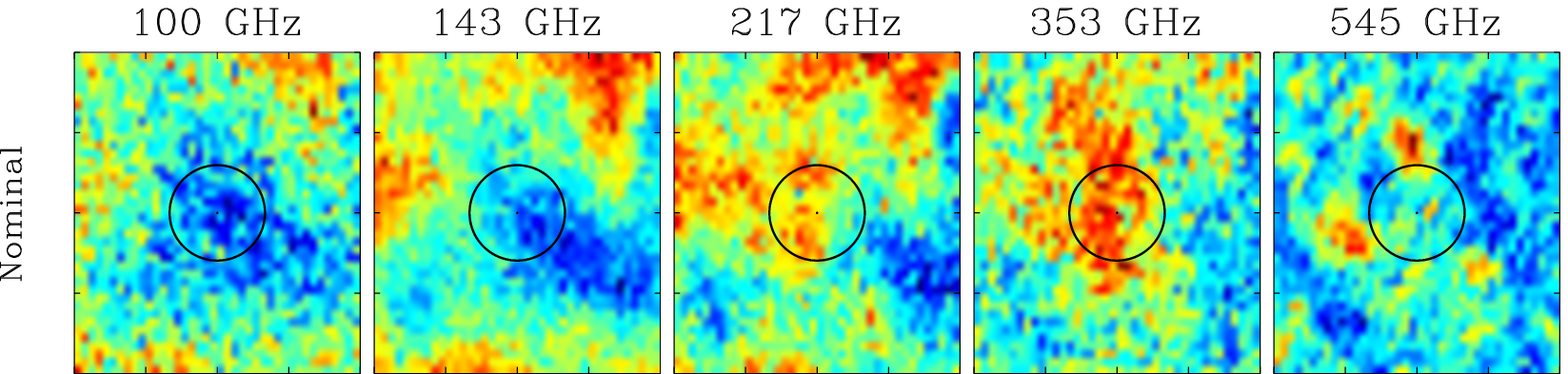}\\
\epsfxsize=1.0\textwidth \epsfbox{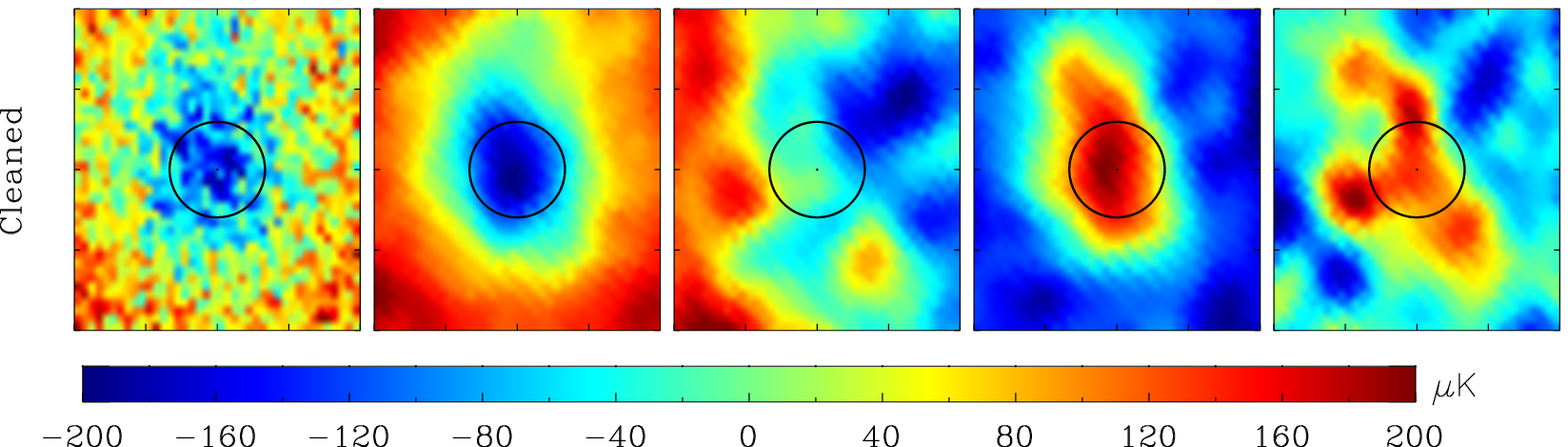}
\caption{First and second rows: 
{\it Planck} Nominal and  foreground cleaned patches
centered at the position of A1656 (Coma cluster)
at 100-545~GHz. Patches are 2\textdegree$\times$2\textdegree.}\label{fig1}
\end{figure*}

At each frequency, from 100 GHz to 545 GHz, our cleaning method provides 
foreground cleaned patches $\mathcal{P}(\nu, x)$ 
centered on the position $x$ of the selected cluster from where 
we can measure the TSZ temperature fluctuations ($\delta \bar{T}/T_0$)
at the cluster position. In Fig. \ref{fig1} we show the patch around the Coma cluster to
illustrate how effectively our cleaning procedure removes foregrounds. 
Temperatures are given in $\mu$K. The top row corresponds to
the original {\it Planck} Nominal data and the bottom row to 
foreground cleaned maps. 
The Nominal patches are dominated by the intrinsic CMB temperature
fluctuations, while  in the foreground clean maps the TSZ signal dominates.
The signal is negative at 100 and 143 GHz,  zero at 217 and positive at 353 GHz, 
as expected. At 545 GHz is dominated by dust residuals and, together with
217 GHz that corresponds to the TSZ null, they will not be
included in our analysis.

The pressure profile is constructed by taking averages on
the central disc of radius $5$ arcminutes and on rings of 5 arcminutes width out 
to 100 arcminutes. The angular scale associated to each temperature average
is the mean angular distance to the center of the cluster of all pixels
within the disc or ring. The dispersion around the mean ranges from a minimum of
$\sim0.5$ arcminutes in the central 15 arcminutes to a maximum of $\sim 1.5$ arcmin.
To compute the error on the measured TSZ anisotropies, we place the cluster 
at 1,000 random positions in the sky. The patch is cleaned
as described before and the mean on the central 5 arcmin disc and 
rings is evaluated. To avoid overlapping with real clusters, we mask an area 
of one degree radius around all clusters in the \cite{Piffaretti2011} catalog. 
We repeat the procedure and compute the correlation matrix ($C_{ij}$) between 
different bins ($\theta_i$) averaging over all random positions
\begin{equation}
C_{ij}(\nu_k)=\frac{\langle [\delta \bar{T}(\nu_k, \theta_i)-\mu(\nu_k, \theta_i)]
[\delta \bar{T}(\nu_k, \theta_j)-\mu(\nu_k, \theta_j)]\rangle}{\sigma(\nu_k, \theta_i)
\sigma(\nu_k, \theta_j)} ,
\label{eq:Cij}
\end{equation}
where $\mu(\nu_k, \theta_i)=\langle\delta \bar{T}(\nu_k, \theta_i)\rangle$, and 
$
\sigma(\nu_k, \theta_i)= 
\langle[\delta\bar{T}(\nu_k, \theta_i)-\mu(\nu_k,\theta_i)]^2\rangle^{1/2}$.
The error bars on the profile of the Coma cluster are the square root of 
the diagonal elements of the correlation matrix. 


\begin{thebibliography}{100}
\bibitem{astier} 
Astier, P., Guy, J., Regnault, N., Pain, R., Aubourg, E., Balam, D., Basa, S., Carlberg, R. G.,
Fabbro, S., Fouchez, D., et al.,  A\&A, {\bf 447}, 31 (2006)

\bibitem{bao:blake} 
Blake, Chris; Kazin, Eyal A.; Beutler, F., Davis, T. M., Parkinson, D., Brough, S., Colless, M., 
Contreras, C., Couch, W., Croom, S., et al., MNRAS, {\bf 418}, 1707 (2011)

\bibitem{clocc} 
Clocchiatti, A., Schmidt, B. P., Filippenko, A.V., Challis, P., 
Coil, A. L., Covarrubias, R., Diercks, A., Garnavich, P., 
Germany, L., Gilliland, R., et al., ApJ, {\bf 642}, 1 (2006)

\bibitem{wmap9} 
Hinshaw, G., Larson, D., Komatsu, E., Spergel, D. N., Bennett, C. L., 
Dunkley, J., Nolta, M. R., Halpern, M., Hill, R. S., Odegard, N., et al.,
ApJS, {\bf 208}, 19H (2013)


\bibitem{planck15_XIII} 
Planck Collaboration. Planck 2015 results. XIII. Cosmological parameters. 
A\&A, submitted. ArXiv:1502.01589v2 (2015)

\bibitem{riess} 
Riess, A.G., Strolger, L.G., Tonry, J., Casertano, S., 
Ferguson, H.C., Mobasher, B., Challis, P., Filippenko, A. V., 
Jha, S., Li, W., ApJ,  {\bf 607}, 655 (2004)

\bibitem{carroll} 
Carroll, S.M., Duvvuri, V., Trodden, M. \& Turner, M.S., Phys Rev D, {\bf 70}, 043528 (2004)

\bibitem{sotiriou} 
Sotiriou, T.P. \& Faroni, V., Rev Mod Phys, {\bf 82}, 451 (2010)

\bibitem{defelice}
De Felice, A. \& Tsujikawa, S., Living Rev Rel, {\bf 13}, 3 (2013)

\bibitem{Nojiri2007}
Nojiri, S.,  Odintsov, S.D.,  Int. \ J.\ Geom.\ Meth.\ Mod.\ Phys.,  {\bf 4}, 115 (2007)

\bibitem{Nojiri2011} 
Nojiri, S.,  Odintsov, S.D., Phys. Rept., {\bf 505}, 59  (2011)

\bibitem{Capozziello2011} 
Capozziello, S., De Laurentis, M.,  Phys. Rept., {\bf 509}, 167 (2011)

\bibitem{demartino2015a} 
de Martino, I., De Laurentis, M., Capozziello, S.,  Universe, {\bf 1}, 123 (2015)

\bibitem{Khoury2004}
Khoury, J., \& Weltman, A., Phys. Rev. D, {\bf 69}, 044026 (2004)

\bibitem{Mota2004}
Mota, D.F., \& Barrow, J.D., Phys. Lett. B, {\bf 581}, 141 (2004)

\bibitem{hu1} 
Ferraro,S., Schimdt,F., Hu,W., Phys Rev  D, {\bf 83}, 063503 (2011)

\bibitem{hu2007-1} 
Hu, W. \& Sawicki, I., Phys Rev D, {\bf 76}, 064004 (2007)

\bibitem{hu2007-2} 
Hu, W. \& Sawicki, I., Phys Rev D, {\bf 76}, 104043 (2007)

\bibitem{Li21015} 
Li, B.,He, J.-H., Gao, L., MNRAS, {\bf 456}, 146 (2015)

\bibitem{lima} 
Lima, N.A. \& Liddle, A.R., Phys Rev D, {\bf 88}, 043521 (2013)

\bibitem{sawicki} 
Sawicki, I. \& Hu, W., Phys Rev D,  {\bf 75}, 127502 (2007)

\bibitem{hu2} 
Schimdt, F., Vikhlinin A., Hu, W., Phys. Rev. D,  {\bf 80}, 083505 (2009)

\bibitem{song} 
Song, Y.S., Hu, W. \& Sawicki, I., Phys Rev D,  {\bf 75}, 044004 (2007)

\bibitem {arnold} 
Arnold, C., Puchwein, E. \& Springer, V., MNRAS, {\bf 440}, 833 (2014)

\bibitem{Terukina2014}
Terukina, A., Lombriser, L., Yamamoto, K., Bacon, D., Koyama, K., \& Nichol, R. C., JCAP, {\bf 04}, 013 (2014)

\bibitem{Terukina2015}	
Terukina, A., Yamamoto, K., Okabe, N., Matsushita, K., Sasaki, T., JCAP, {\bf 10}, 64 (2015)


\bibitem{Wilcox2015} 	
Wilcox, H., Bacon, D., Nichol, R. C., Rooney, P. J., Terukina, A., Romer, A. K.,
Koyama, K., Zhao, G.-B., Hood, R., Mann, R. G. et al., MNRAS, {\bf 452}, 1171 (2015)

\bibitem{CapFar2010}
Capozziello, S. \& Faraoni, V., Beyond Einstein Gravity: A Survey of 
gravitational Theories for Cosmology and Astrophysics. Springer, Heidelberg (2010)

\bibitem{stabile}  
Capozziello, S. Stabile, A. \& Troisi, A., Modern Physics Letters A, {\bf 24}, 09 (2009)

\bibitem{annalen} 
Capozziello, S. \& De Laurentis, M., Ann. Phys., {\bf 524}, 1 (2012)

\bibitem{hans} 
Quandt, I. \& Schmidt, H. J., Astron. Nachr., 312, {\bf 97} (1991)

\bibitem{tartaglia}  
Allemandi, G., Francaviglia, M., Ruggiero, M. and Tartaglia A., Gen. Rel. Grav., {\bf 37}, 1891 (2005)


\bibitem{berry} 
Berry, C. P. L. \& Gair, J. R., Phys Rev D, {\bf 83}, 104022 (2011)

\bibitem{trosi}  
Capozziello, S. \& Troisi, A,  Phys  Rev  D, {\bf 72}, 044022 (2005)

\bibitem{tsuji} 
Capozziello, S. \& Tsujikawa, S., Phys  Rev  D,  {\bf 77},  107501 (2008)

\bibitem{jeans}  
Capozziello, S., De Laurentis, M., De Martino, I., Formisano, M., Odintsov, S. D.,  Phys  Rev  D, {\bf 85}, 044022 (2012)


\bibitem{delaurentis2013}
De Laurentis, M., De Martino, I., MNRAS, {\bf 43}, 741 (2013)

\bibitem{delaurentis2015}
De Laurentis, M., De Martino, I., Int. J. Geom. Methods Mod. Phys. {\bf 12}, 1550040 (2015)

\bibitem{cardone} 
Cardone, V. F. \& Capozziello, S., MNRAS, {\bf 414}, 1301 (2011)

\bibitem{napolitano} 
Napolitano, N. R., Capozziello, S., Romanowsky, A. J., Capaccioli, M. \& Tortora, C., 
ApJ, {\bf 748}, 87 (2012)

\bibitem{salzano} 
Capozziello, S., De Filippis, E. \& Salzano, V., MNRAS, {\bf 394}, 947 (2009)

\bibitem{tsz}
Sunyaev, R.~A., \& Zeldovich, Y.~B., Comments on Astrophys. Space Phys., {\bf 4}, 173 (1972)

\bibitem{demartino2014}
De Martino, I.; De Laurentis, M.; Atrio-Barandela, F.; Capozziello, S., MNRAS, {\bf 442}, 921 (2014)

\bibitem{veritas2012} 
The Veritas Collaboration, ApJ, {\bf 757}, 123  (2012)

\bibitem{Stelle78}
Stelle, K., Gen. Relat. Gravit., {\bf 9}, 343 (1978)

\bibitem{weinberg}
Weinberg, S.,Gravitation and Cosmology. New York,Wiley  (1972)

\bibitem{eingorn}
Eingorn, M., \& Zhuk, A., Phys. Rev. D, {\bf 84}, 024023 (2011)

\bibitem{CapNoj2006_1}
Capozziello, S., Nojiri, S.  and Odintsov, S.D., Phys. Lett.B 634, {\bf 93} (2006)

\bibitem{CapNoj2006_2}
Capozziello, S., Nojiri, S., Odintsov, S.D. and  Troisi, A., Phys. Lett.B, {\bf 639}, 135 (2006)

\bibitem{Nojiri2006}
Nojiri, S.,  Odintsov, S.D., Phys. Rev. D, {\bf 74}, 086005 (2006)

\bibitem{cap2006}  
Capozziello, S. Stabile, A. \& Troisi, A., Mod. Phys. Lett.A {\bf 21}, 2291 (2006)

\bibitem{sotiriou2006} 
Sotiriou, T.P., Class. Quant. Grav.  {\bf 23}, 5117 (2006)

\bibitem{cap2008}
Capozziello, S., Corda, C., \& de Laurentis, M., Physics Letters B, {\bf 669}, 255 (2008)

\bibitem{ksz} 
Sunyaev, R. A. \& Zeldovich, Y. B., MNRAS, {\bf 190}, 413 (1980)

\bibitem{fixsen}
Fixsen, D.J., ApJ, {\bf 707}, 916 (2009)

\bibitem{Vikhlinin2006}
Vikhlinin, A., Kravtsov, A., Forman, W., et al., ApJ, {\bf 640}, 691 (2006)

\bibitem{itoh1998} 
Itoh, N., Kohyama, Y. \& Nozawa, S., ApJ, {\bf 502}, 7  (1998)

\bibitem{nozawa1998} 
Nozawa, S., Itoh, N. \& Kohyama, Y., ApJ, {\bf 508}, 17 (1998)

\bibitem{nozawa2006} 
Nozawa, S., Itoh, N., Suda, Y. \& Ohhata, Y., Nuovo Cimento, {\bf 121}, 487 (2006)

\bibitem{gorski2005} 
Gorski, K., Hivon, E., Banday, A., Wandelt, B.D., Hansen, F.K., Reinecke, M. \& Bartelmann, M., ApJ, {\bf 622}, 759 (2005)

\bibitem{Reiprich2002} 
Reiprich, T. H.; B\"{o}hringer, H., ApJ,  {\bf 567}, 716 (2002)

\bibitem{PLANCKX2012} 
 Planck Collaboration. Planck Intermediate Results X: Physics of the hot gas in the Coma cluster, 
 A\&A, {\bf 554}, 140 (2013)
 

\bibitem{Coma_xray3}
Simionescu, A., Werner, N., Urban, O., et al. 
ApJ, {\bf 775}, 4  (2013)

\bibitem{Coma_xray4}
Sato, T., Matsushita, K., Ota, N., Sato, K., Nakazawa, K., Sarazin, C.L.,
Publ. Astron. Soc. Jpn. {\bf 63} (2011) 991

\bibitem{Coma_xray5}
 Neumann,  D.M., Lumb,  D.H.,  Pratt, G.W. and  Briel, U.G.
  A\&A {\bf 400}, 811, (2003)

\bibitem{Coma_xray6}
 Sanders, J.S., et al., Science, {\bf 341},  6152 (2013)

\bibitem{Coma_SB}
Churazov, E., Vikhlinin, A., Zhuravleva, I., Schekochihin, A., Parrish, I., Sunyaev, R., Forman, W., B\"{o}hringer, H., Randall, S., 
MNRAS, {\bf 421}, 1123, (2012) 

\bibitem{Coma_XT1}
Snowden, S. L., Mushotzky, R. F., Kuntz, K. D., Davis, D. S.
A\&A {\bf 478} (2008) 615.

\bibitem{Coma_XT2}
Wik, D.R. et al.,
ApJ {\bf 696}, 1700, (2009).

\bibitem{Gastaldello}
Gastaldello, F. et al., ApJ, {\bf 800}, 139, (2015)

\bibitem{FuscoFemiano}
Fusco-Femiano, R., Lapi, A., Cavaliere, A., 
ApJ, {\bf 763}, 1, (2013)

\bibitem{demartino2015b} 
de Martino, I., G\'{e}nova-Santos, R., Atrio-Barandela, F., Ebeling, H.,  
Kashlinsky, A., Kocevski,  D. \& Martins,  C.J.A.P., ApJ, {\bf 808}, 128 (2015)
 
 
\bibitem{Hastings1970} 
Hastings, W.K., Biometrika, {\bf 57}, 97 (1970)
 
\bibitem{Metropolis1953} 
Metropolis, N., et al., J. Chem., Phys., {\bf 21}, 1087 (1953)

\bibitem{Gelman1992} 
Gelman, A., and Rubin, D.B., Statist. Sci., {\bf 7}, 457 (1992)

\bibitem{Gelman1996}  
Gelman, A., Roberts, G. O.,  Gilks, W. R., Bayesian statistics, {\bf 5}, 599 (1996)

\bibitem{Roberts1997}  
Roberts, G. O.,  Gelman, A., Gilks, W. R., Ann. Appl. Probab.,  {\bf 7}, no. 1, 110 (1997)

\bibitem{Sarazin1988}
Sarazin, C.L. (1988), "X-ray Emission from Clusters of Galaxies", 
Cambridge Astrophysics Series, Cambridge University Press.


\bibitem{Verde2003} 
Verde, L., Peiris, H. V., Spergel, D. N., Nolta, M. R., Bennett, C. L.,
Halpern, M., Hinshaw, G., Jarosik, N., Kogut, A., Limon, M., et al.,
ApJS, {\bf 148}, 195 (2003)


\bibitem{Ascasibar2006} 
Ascasibar, Y.; Sevilla, R.; Yepes, G.; M\"{u}ller, V.; Gottl\"{o}ber, S., MNRAS, {\bf 371}, 193 (2006)

\bibitem{Bode2009}
Bode, P., Ostriker, J. P., Vikhlinin, A., ApJ, {\bf 700}, 989 (2009)

\bibitem{capelo2012}
Capelo, P. R.; Coppi, P. S.; Natarajan, P., MNRAS, {\bf 422}, 686C (2012)

\bibitem{Ostriker2005} 
Ostriker, J. P., Bode, P., \& Babul, A., ApJ, {\bf 634}, 964 (2005)

\bibitem{Bertschinger1985}
Bertschinger, E., ApJS, {\bf 58}, 39 (1985)



\bibitem{Brax2015}
Brax, P., Rizzo, L.A., and Valageas, P., Phys. Rev. D, {\bf 92}, 043519 (2015)


\bibitem{bernal2015} 
Bernal, T., L\'{o}pez-Corona, O., Mendoza, S.,  arXiv:1505.00037 (2015)
 
 
\bibitem{bobin2013}
Bobin, J., Sureau, F., Paykari, P., Rassat, A., Basak, S., \& Starck, J.-L., A\&A, {\bf 553}, L4 (2013)

\bibitem{bobin2014} 
Bobin, J., Sureau, F., Starck, J.-L., Rassat, A. \& Paykari, P., A\&A, {\bf 563}, 105 (2014)

\bibitem{planck13_XIII} 
Planck Collaboration, "Planck 2013 results. XIII. Galactic CO emission", A\&A, {\bf 571}, 13 (2014)

\bibitem{planck13_XVIII} 
Planck Collaboration,   "Planck 2013 results. XXVIII. The Planck Catalogue of Compact Sources", A\&A, {\bf 571}, 28 (2014)

\bibitem{planck13_XXIX} 
Planck Collaboration, "Planck 2013 results. XXIX. 
The Planck catalogue of Sunyaev-Zeldovich sources", A\&A, {\bf 571}, 29 (2014)

\bibitem{planck13_XI} 
Planck Collaboration, "Planck 2013 results. XI. All-sky model of thermal dust emission", A\&A, {\bf 571}, A11 (2014)

\bibitem{planck13_XII} 
Planck Collaboration,  "Planck 2013 results. XII. Component separation", A\&A, {\bf 571}, 12 (2014)

\bibitem{diego2002}
Diego, J.M., Vielva, P., Mart\'{i}nez-Gonzalez, E., Silk, J. \& Sanz, J.L., MNRAS, {\bf 336}, 1351 (2002) 
 
\bibitem{Piffaretti2011} 
Piffaretti R., Arnaud M., Pratt G.W., Pointecouteau E., Melin J.-B., A\&A,  {\bf 534}, 109  (2011)

\bibitem{PRISM}
PRISM Collaboration; Andr\'e P. et al., JCAP, {\bf 1402}, 006 (2014)
  
\end{thebibliography}
\end{document}